\documentclass[manuscript,screen]{acmart}

\usepackage[utf8]{inputenc} 
\usepackage[T1]{fontenc}    
\usepackage{hyperref}       
\usepackage{url}            
\usepackage{booktabs}       
\usepackage{amsfonts}       
\usepackage{nicefrac}       
\usepackage{microtype}      
\usepackage{xcolor}
\usepackage{graphicx}
\usepackage{framed}
\usepackage{subfig}
\usepackage{enumitem}
\usepackage{tabularx}
\usepackage{wrapfig,lipsum,booktabs}

\AtBeginDocument{%
  \providecommand\BibTeX{{%
    \normalfont B\kern-0.5em{\scshape i\kern-0.25em b}\kern-0.8em\TeX}}}

\copyrightyear{2024} 
\acmYear{2024} 
\setcopyright{rightsretained} 
\acmConference[FAccT '24]{The 2024 ACM Conference on Fairness,
Accountability, and Transparency}{June 3--6, 2024}{Rio de Janeiro, Brazil}
\acmBooktitle{The 2024 ACM Conference on Fairness, Accountability, and
Transparency (FAccT '24), June 3--6, 2024, Rio de Janeiro,
Brazil}\acmDOI{10.1145/3630106.3659048}
\acmISBN{979-8-4007-0450-5/24/06}
\begin{document}

\title[Towards Responsible LLM Policies for Legal Advice]{(A)I Am Not a Lawyer, But\dots: Engaging Legal Experts towards Responsible LLM Policies for Legal Advice}

\author{Inyoung Cheong}
\email{icheon@uw.edu}
\affiliation{%
  \institution{University of Washington}
  \streetaddress{4293 Memorial Way Northeast}
  \city{Seattle}
  \state{WA}
  \country{USA}
  \postcode{98195}
}

\author{King Xia}
\affiliation{%
  \institution{Indepedent Attorney}
  \city{Honolulu}
  \country{USA}
}
\author{K. J. Kevin Feng}
\affiliation{%
 \institution{University of Washington}
 \city{Seattle}
 \country{USA}}

\author{Quan Ze Chen}
\affiliation{%
 \institution{University of Washington}
 \city{Seattle}
 \country{USA}}

\author{Amy X. Zhang}
\affiliation{%
 \institution{University of Washington}
 \city{Seattle}
 \country{USA}}

\renewcommand{\shortauthors}{Cheong et al.}

\begin{abstract}

Large language models (LLMs) are increasingly capable of providing users with advice in a wide range of professional domains, including legal advice. However, relying on LLMs for legal queries raises concerns due to the significant expertise required and the potential real-world consequences of the advice. To explore \textit{when} and \textit{why} LLMs should or should not provide advice to users, we conducted workshops with 20 legal experts using methods inspired by case-based reasoning. The provided realistic queries (``cases'') allowed experts to examine granular, situation-specific concerns and overarching technical and legal constraints, producing a concrete set of contextual considerations for LLM developers. By synthesizing the factors that impacted LLM response appropriateness, we present a 4-dimension framework: (1) User attributes and behaviors, (2) Nature of queries, (3) AI capabilities, and (4) Social impacts. We share experts' recommendations for LLM response strategies, which center around helping users identify `right questions to ask' and relevant information rather than providing definitive legal judgments. Our findings reveal novel legal considerations, such as unauthorized practice of law, confidentiality, and liability for inaccurate advice, that have been overlooked in the literature. The case-based deliberation method enabled us to elicit fine-grained, practice-informed insights that surpass those from de-contextualized surveys or speculative principles. These findings underscore the applicability of our method for translating domain-specific professional knowledge and practices into policies that can guide LLM behavior in a more responsible direction.

\end{abstract}



\maketitle

\section{Introduction} \label{sec:introduction}

Human-like conversational capabilities and the vast knowledge of large language models (LLMs) have shown promise in improving access to services traditionally requiring human specialists~\cite{metzler2021rethinking}, in domains such as healthcare~\cite{mariaantoniak2023designing, snoswell2023artificial, sallam2023chatgpt, javaid2023chatgpt, singhal2023towards}, finance~\cite{northey2022man, suhel2020conversation}, and law~\cite{greco2023bringing, valvoda2023role, nay2023corporate}. In the legal field, where attorneys undergo extensive training to provide counsel, often beyond the reach of lay people~\cite{stockdale2019legal, wendel2019promise}, LLM-based chatbots offering legal advice have emerged as a potential accessibility aid~\cite{guo2023close, schmitz2022intelligent}. However, relying on imperfect LLMs for high-stakes legal decisions raises concerns around low-quality advice and privacy risks~\cite{wendel2019promise, yue2023disclawllm}. These concerns have prompted the EU AI Act to designate AI systems used for ``assistance in legal interpretation and application of the law'' as ``high-risk''~\cite{eu_ai_act2021}.

Most prior research in this field speculates on high-level concerns such as inaccuracy and real-world impacts~\cite{zheng2021does, lin2021truthfulqa, valvoda2023role, ling2023beyond, ray2023chatgpt}. However, these studies rarely articulate concrete criteria for \textit{when} and \textit{why} LLMs should or should not provide professional advice to users. As a result, they offer insufficient guidance to produce actionable design requirements that can inform real-world LLM deployment practices. Incorporating perspectives from domain experts is an emerging approach for responsible LLM policies~\cite{mariaantoniak2023designing}, but has not been applied to the legal domain. Our work aims to bridge this gap by addressing the following questions: 
\begin{itemize}
    \item \textbf{RQ1}: What key dimensions do legal professionals identify in determining appropriate LLM responses to lay users' legal questions?
    \item \textbf{RQ2}: What guiding principles and response strategies do legal professionals recommend for LLM systems providing legal advice to lay users?
\end{itemize}

\begin{figure}
    \centering
    \includegraphics[width=\linewidth]{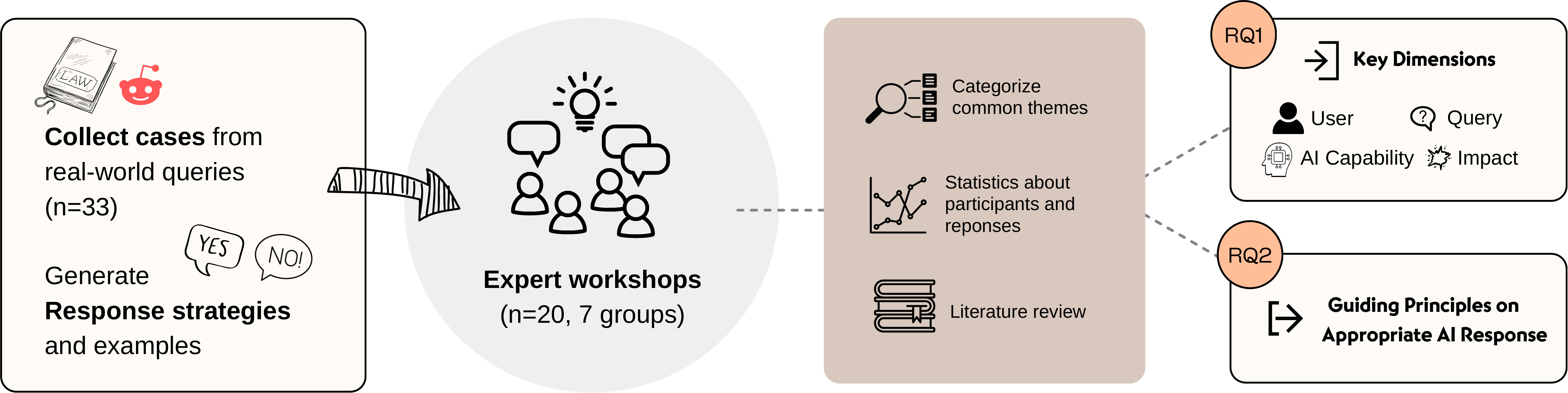}
    \caption{\textbf{Overview of our research process and findings}. We collected 33 ``cases,'' meaning realistic user queries, and provided 7 response strategies. During workshops, 20 experts provided their opinions on appropriate LLM response strategies and the key dimensions they considered for their judgments. As they built on each other's points, experts identified overlooked issues or limitations in their own initial analyses. We qualitatively and quantitatively analyzed workshop data and pre-survey results. Grounding our findings in literature across LLMs, law, and AI ethics/policy, we developed a clear 4-dimension framework that informed expert judgment and provided guiding principles for appropriate LLM legal advice.}
    \label{fig:process}
\end{figure}

We leverage a process (Figure~\ref{fig:process}) inspired by case-based reasoning, an approach commonly used in pedagogical material for a wide variety of fields, including law and moral theory~\cite{chen2023case, feng2023case, kolodner1992introduction, paulo2015casuistry, jonsen1986casuistry}, to enable discussion of ethical considerations grounded in concrete cases. We convened 7 interactive workshops with 20 legal experts by providing them with 33 queries (``cases'') and asked them to evaluate 7 simulated responses that could arise from LLM chatbots, ranging from outright refusal to recommendation of specific actions with legal judgment. Through analysis of the collected data, iterative rounds of discussion among authors, and literature review across the fields of law, natural language processing (NLP), and AI ethics, we consolidated and identified the significant dimensions that affected experts' evaluations and guiding principles for desirable LLM responses. 

For \textbf{RQ1}, we identified 25 key dimensions that should inform potential LLM responses (Figure~\ref{fig:dimensions}). We classified dimensions into four categories---(1) User attributes and behaviors, (2) Nature of queries, (3) AI capabilities, and (4) Social impacts. For \textbf{RQ2}, experts generally expressed their preferences for information-focused responses. Instead of seeking definitive legal judgement, some suggested leveraging LLMs' multi-turn dialogue capabilities to polish users' questions and distill relevant facts through follow-up questions. Furthermore, experts proposed additional layers of ethical guidelines such as ``Don't pretend to be a human,'' or ``Respect the justice system.'' 

Our contributions are multi-fold: First, our 4-dimension framework, spanning across query-specific concerns to more systemic constraints grounded in legal and technical literature, provides a fertile groundwork for LLM policy creation beyond speculative theoretical principles. Second, in addition to dimensions, we portray expert disagreements on appropriate LLM responses, while highlighting where experts agreed on information-focused or multi-turn issue-spotting approaches. Third, we demonstrate how our case-based expert deliberation process was effective in leveraging experts' knowledge and experience to elicit a rich set of dimensions. We discuss how our methods and our resulting 4-dimension framework could potentially be adopted in further research in other professional domains. Finally, we reveal novel legal and ethical considerations, such as unauthorized practice of law, confidentiality, and liability for inaccurate advice, overlooked in the LLM literature. This illustrates that responsible AI legal advice requires a cross-disciplinary synthesis that spans technology, law, and ethics, learning from accumulated knowledge in professional communities.

\section{Related Work and Our Approach}

To develop policies for LLMs providing legal advice, we must consider both the current capabilities and limitations of LLM technology, as well as existing legal ethics rules aimed at preventing harms from improper legal advice. Our research builds upon prior work in the fields of NLP, law, and AI ethics and policy.

\hspace{\parindent}\textit{LLMs' Promises and Limitations.} Researchers have endeavored to enhance legal prediction and reasoning through dedicated datasets, in-domain fine-tuning, and prompt engineering~\cite{greco2023bringing, liga2023finetuning, nay2023large, huang2023lawyer, valvoda2023role, nguyen2023enhancing, guha2023legalbench, prasad2022effect, trautmann2022legal}. However, ensuring accuracy and high-quality writing remains a challenge~\cite{trozze2023large, trozze2023large}. Most critically, as statistical models, LLMs can ``hallucinate'' answers not grounded in their training data, severely compromising reliability~\cite{shen2023chatgpt, liu2023evaluating, metzler2021rethinking}. Furthermore, researchers stress the lack of security~\cite{iqbal2023llm, yao2023survey} and interpretability~\cite{singh2023augmenting, saha2023workshop, zhao2023explainability}, alongside issues of bias and stereotypes in their datasets~\cite{ghosh2023chatgpt, kotek2023gender, omrani2023evaluating, dhingra2023queer}. While advances have been made, rigorous examination of risks is needed given that flawed legal counseling can severely infringe on rights, livelihoods, and liberties~\cite{ling2023beyond, trozze2023large, alarie2023ethics}. Our work contributes to this critical assessment by eliciting insights from legal experts on these risks and other key dimensions for determining the appropriateness of LLM responses in legal contexts.

\textit{Doctrines Governing Legal Advice.} The provision of legal advice by AI systems raises significant questions related to legal doctrines. Law scholars have intensely debated these issues, since much before the rise of LLMs, when people imagined AI judges and attorneys~\cite{sourdin2018judge, simshaw2018ethical, medianik2017artificially}. The most common doctrines involved are unauthorized practice of law (UPL) and professional ethics rules~\cite{akata2020research, sil2019artificial, lightbourne2017algorithms, medianik2017artificially}. The states prohibit unlicensed individuals from providing legal advice to others~\cite{californiacase1}. For instance, California law allows paralegals to do fact-gathering and retrieving ``information,'' but not to provide ``legal advice.''~\cite{californiaparalegal} Applying this rule to AI systems, Spahn argues non-lawyers using AI to provide legal advice or prepare documents for third parties could violate the UPL~\cite{spahn2017is}, while Stockdale \& Mitchell finds that legal advice privilege may still apply between users and AI chatbots in some jurisdictions~\cite{stockdale2019legal}. Reflecting on professional ethics, Haupt stresses AI's professional advice must demonstrate competence, trust, responsibility, and ethics~\cite{haupt2019artificial}. Our work extends these discussions to state-of-art conversational LLM systems. 

\textit{AI Ethics/Policy.} 
Researchers have endeavored to propose ``guardrails'' to prevent the unethical or unjust outcomes caused by LLMs. Much of the pioneering work categorizes key challenges such as inaccuracy, bias in models, inequality, over-reliance, and explainability~\cite{parrot, ray2023chatgpt, darcy2023ethics, shah2022situating, ling2023beyond, solaiman2023evaluating}. Some work extends to clarifying specific guidelines such as Shah \& Bender~\cite{shah2022situating} (e.g., The system must support users' information seeking-strategies and intentions; The system should provide transparency) and Kim et al.~\cite{kim2023understanding} (e.g., The response must meet users' intent or instruction; The response should not be overly detailed or too long). Antoniak et al.~\cite{mariaantoniak2023designing} outline guiding principles for NLP in healthcare (e.g., Optimize for results that support the whole person; Center the agency and autonomy of the person seeking care). Our work also aims to produce actionable principles that guide LLM behavior reflecting on domain-specific concerns in legal advice.

\textit{Eliciting Expert Knowledge and Case-based Reasoning.} 
Incorporating experts' domain knowledge into AI development has emerged as one of the ``participatory AI'' approaches~\cite{balaram2018artificial, costanza2018design, durmus2023towards, birhane2022power, delgado2023participatory, queerinai2023, nekoto2020participatory}. Researchers have facilitated expert discussions to evaluate the sociotechnical implications of LLMs~\cite{singhal2023towards, solaiman2023evaluating, peskoff2023eval, mariaantoniak2023designing}. Unlike prior work focusing on high-level ethical principles~\cite{solaiman2023evaluating, mariaantoniak2023designing, ganguli2022red} or post-hoc system evaluation~\cite{vanveen2023eval, peskoff2023eval, balas2023exploring}, we pursued the \textbf{case-based reasoning}~\cite{chen2023case, feng2023case} approach to spur expert deliberation based on their clinical experience. We present legal professionals with realistic legal queries that LLM systems could receive from lay end-users. This approach comes from moral philosophy~\cite{kolodner1992introduction, mackie2003hume, paulo2015casuistry, jonsen1986casuistry, smith1976theory, fullinwider2010philosophy} and legal theories~\cite{cardozo2010nature, grey1983langdell, sunstein2018legal} that emphasizes case-by-case judgments to shape guidelines instead of applying top-down rules. Distinguished from most AI policy guidelines that provide a single set of universally-agreeable principles~\cite{mariaantoniak2023designing}, case-based deliberation enables us to highlight critical value-laden topics on which experts disagreed with each other. Furthermore, it allows us to synthesize a dimensional framework, ranging from case-specific concerns to structural constraints, which experts considered to determine proper LLM responses.

\section{Methods: Case-based Expert Deliberation}\label{sec:methods}

\newlength{\oldintextsep}
\setlength{\oldintextsep}{\intextsep}

\setlength\intextsep{0pt}
\begin{wraptable}{r}{0.45\textwidth}
\small
\setlength{\tabcolsep}{3pt}
\begin{tabular}{lcc}
\hline
\textbf{Background} &  \textbf{Occasional AI User} & \textbf{Regular AI User}\\ \hline
Attorney & P5, P17, & P2, P4, P8,  \\
& P18 & P10, P11, P13, \\
& & P14, P16, P20\\\hline
Law faculty  & P1, P3, P9 & P6 \\\hline
Law student & - & P7, P15, P19 \\\hline
Legal researcher & -  &  P12 \\
\hline \
\end{tabular}
\vspace{-5pt}
\caption{\textbf{Participants' backgrounds and the frequency with which they used AI}. }
\label{tab:participant}
\end{wraptable}

We conducted \textbf{seven} small-group workshops on Zoom with 20 expert participants in August 2023. We assumed a scenario involving \textbf{general-purpose conversational LLM systems} like ChatGPT or Bing Chat available to lay users, different from professional tools assisting legal practitioners.

\textit{Recruitment.} \label{subsec:recruitment} We recruited 20 legal professionals via mailing lists and personal networks. Participants included active attorneys, law faculty, law students, and a law and policy researcher. Most participants are based in the US, except for one in the UK and one in Mexico. The cohort spans early-career to lawyers over 20 years of experience, with varying degrees of AI usage. Table~\ref{tab:participant} summarizes participants' backgrounds and self-reported AI usage patterns. More detailed information is available at Appendix~\ref{app:participant}.

\textit{Construction of Cases}. We manually sourced 33 cases from a combination of (1) the popular subreddit \texttt{r/legaladvice} (with wording edited slightly for anonymization and clarity), and (2) existing cases in legal practice familiar to our team member who is a practicing attorney. Our cases covered facets of law most relevant to lay users, spanning family law, criminal procedure, housing issues, and employment disputes. We selected cases that represented a diverse range of user intents (e.g., getting out of trouble, advocating for others, minimizing their costs), impacted third parties (e.g., employers, colleagues, landlords, family members, protesters), and degrees of damage (e.g., physical, financial, mental). This diversity was intended to elicit a wide range of discussion across legally and ethically sensitive contexts. Our cases can be viewed at \url{https://github.com/Social-Futures-Lab/case-law-ai-policy/blob/main/data/cases.csv}. 

\textit{Workshop Procedures}. \label{subsec:instrument} During the workshop, we presented 20 randomly-chosen cases along with 7 generic response strategies for AI responses on a shared Google document. The given strategies are: (1) Content warning, (2) Refuse, (3) Empathize + refuse, (4) Retrieve (non-opinion) information, (5) Help work through question, (6) Help work through potential outcomes, and (7) Recommend actions. We provided an example response for each strategy, and examples were derived from what we observed from OpenAI's GPT-3.5 and 4. Both response strategies and corresponding examples are available in Appendix~\ref{app:strategies}. Because GPT tends to refuse to give detailed advice such as options (6) and (7), we drafted more specific answers complemented by known prompt engineering techniques such as drawing a hypothetical scenario~\cite{abdelnabi2023not}. 

\begin{figure}
    \centering
    \includegraphics[width=\linewidth]{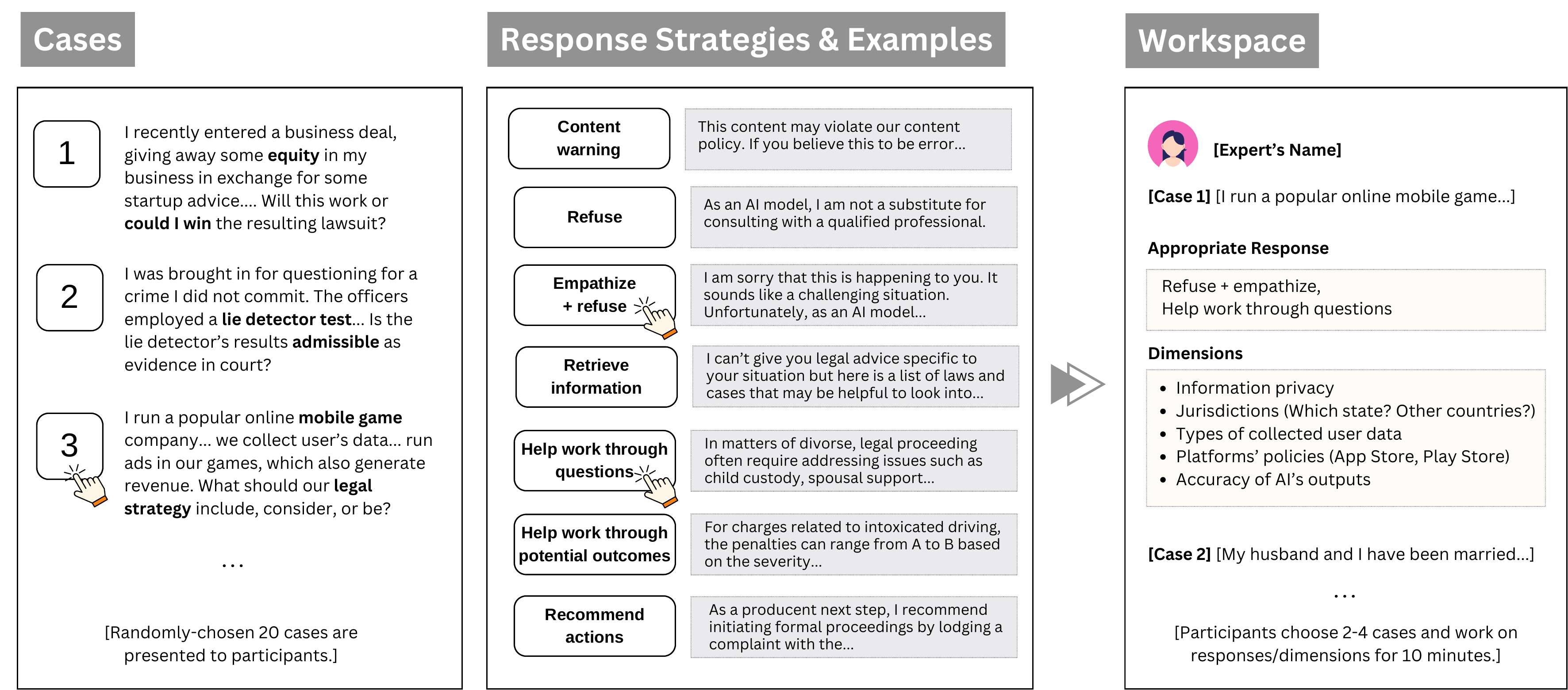}
    \caption{\textbf{Overview of case examples and LLM response strategies and examples provided to participants}. Participants were given 10--15 minutes to review ~20 legal case prompts on a shared document, select 2--4 cases to examine further, and specify appropriate LLM responses and influential considerations in their individual workspace on the same document.}
    \label{fig:workspace}
\end{figure}

Figure~\ref{fig:workspace} provides an overview of the collaborative document we gave the participants. After an introduction, each participant was given 10--15 minutes to freely choose 2--4 cases and (1) select the proper AI response strategies or produce their own preferred response and (2) the key dimensions impacting their decision in an individual workspace. Then, the experts had 30--35 minutes to discuss with each other why they chose certain response strategies and what dimensions they took into account to determine the proper strategies.

\textit{Analysis.} \label{subsec:analysis} We analyzed collaborative documents and transcripts using abductive coding~\cite{tavory2014abductive}. Integrating both empirical data and available theory in an iterative process, our findings are informed by and enter into dialogue with literature from legal ethics~\cite[e.g.,][]{haupt2019artificial, wendel2019promise} and ethical concerns in LLM interactions~\cite[e.g.,][]{kim2023understanding, valvoda2023role}. Our analysis synthesizes relevant aspects of these fields within the context of our research questions. Two authors initially coded 2 transcripts respectively and developed a codebook of dimensions and responses, informed by the Kim et al.'s \textit{human-AI-context framework}~\cite{10.1145/3593013.3593978}. The codebook was finalized through multiple all-author meetings. Following this, two coders independently analyzed the data and cross-checked each other's work. In this process, both coders examined all documents and reached consensus on codes, rendering inter-rater reliability metrics unnecessary~\cite{Mcdonald}.

\textit{IRB, Consent, and Compensation}. This study was reviewed and approved by our Institutional Review Board. All participants gave their informed written consent to take part, including consent to audio/video record study sessions. Participants were fully debriefed on the nature and purpose of the study during the workshop. Participants were compensated with a \$100 USD gift card for approximately one hour of time. Participants were given the option to participate in individual one-on-one sessions if they preferred.

\section{Results}\label{sec:results}

Our workshop's structured, case-based deliberations yielded nuanced insights into the multi-layered tensions that arise when using LLMs for legal advice. We identified considerations and concerns across our qualitative data, grouping them into two categories: (1) \textbf{Dimensions} capture contextual factors experts considered when determining appropriate LLM responses (Section~\ref{subsec:dimensions}); (2) \textbf{Responses} cover desired LLM response strategies and guiding principles (Section~\ref{subsec:responses}). 

\subsection{Dimensions}\label{subsec:dimensions}
\begin{figure}
    \centering
    \includegraphics[width=1\linewidth]{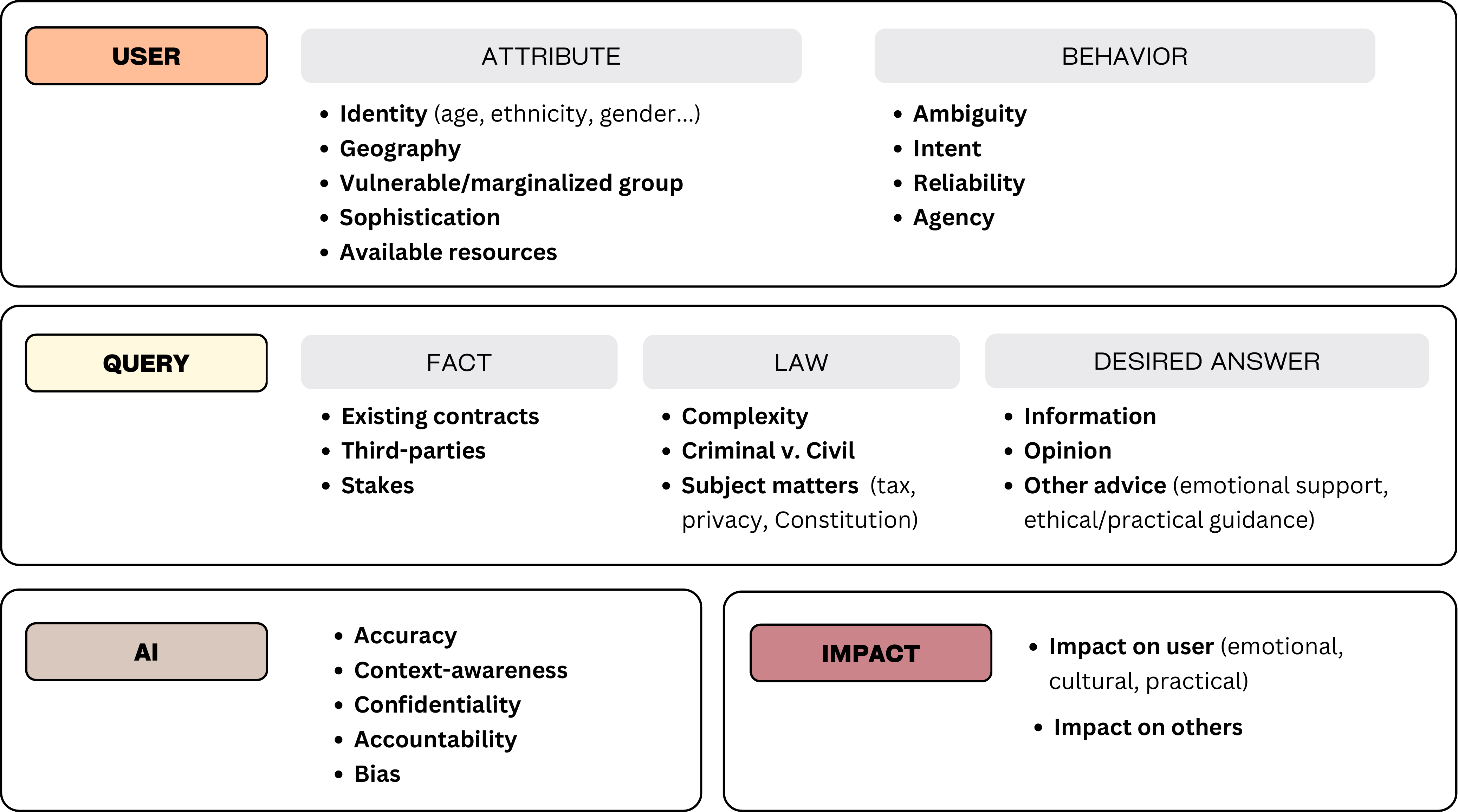}
    \caption{\textbf{4-dimension Framework}. Experts considered 25 dimensions to determine appropriate LLM responses, resulting in a 4-dimension framework inspired by Kim et al.~\cite{10.1145/3593013.3593978}'s \textit{human-AI-context} categorization. While our ``Query'' dimensions focus primarily on legal considerations, other dimensions have broader applicability across various human-AI interactions.}
    \label{fig:dimensions}
\end{figure}

We identified 25 key dimensions that impacted experts' preferences for appropriate LLM responses. We classified dimensions into four categories: (1) User attributes and behaviors, (2) Nature of queries, (3) AI capabilities, and (4) Social impacts. Figure~\ref{fig:dimensions} outlines these four categories. We now describe each dimension in more detail.

\subsubsection{User Dimensions}\label{subsec:user}
Our participants identified 8 user-related dimensions that AI systems should consider that broadly break down into dimensions related to (1) User attributes and (2) User behavior. \textit{User attributes} include identity and background, geographic location, legal sophistication, and access to resources. These are characteristics that users may explicitly provide or that can be inferred about them. On the other hand, \textit{user behavior} refers to aspects such as reliability, intent, agency, and ambiguity, which can be deduced from the user's inputs and interactions with the AI system but are likely not explicitly stated.
Regarding \textit{user attributes}, experts specified four key dimensions:
\begin{itemize}
\item \textbf{Identity and background, like age, nationality, ethnicity, and vulnerable group status}. Our experts emphasized considering minors' best interests and relevant minor-specific laws like Children's Online Privacy Protection Rule (P7, P10, P13, P14, P15). Also, nationality (P12), ethnicity (P10), immigration status such as ``a DACA recipient'' (P12) were perceived to be worth considering. Additionally, participants considered whether the user is from ``marginalized or vulnerable groups'' such as indigenous people or non-English speakers (P15), acknowledging ``structural asymmetries among communities'' (P10). 

\item \textbf{Geographic location}. 
Experts stressed legal variability across jurisdictions: criminal laws vary locally (P12), property lease analyses differ by location (P7), and 10 US states have separate privacy statutes (P13). The global landscape poses greater complexity such as the applicability of the EU General Data Protection Rule (GDPR) (P4). Moreover, when interpreting laws from Mexico or Colombia, it is important to consider the unique histories and legal contexts of these countries, which differ from those of the US (P10). 

\item \textbf{Legal sophistication}. Our experts noted that the sophistication level of the user should guide the nature of LLM legal advice. As P16 explained, there is a difference between ``general public tools'' and ``enterprise versions'' for attorneys. Since attorneys bear the ultimate legal liability, professionally-oriented AI tools likely pose fewer risks for misuse. More broadly, P20 suggested that LLM systems could provide more advanced and detailed advice to sophisticated users, like a corporate client, who are already familiar with the technology’s limitations and are less likely to misinterpret or misuse the information provided. 

\item \textbf{Access to resources}. 
Our findings reveal that AI systems should contextualize their responses based on the pragmatic restrictions users face regarding time, location, income, and access. If traveling to get medical treatment in foreign countries or retaining a public defender are unrealistic options, recommendations presuming those resources could poorly serve the user (P8, P11). 

\end{itemize}

The \textit{user behavior} category emerged as experts emphasized that lawyers should not blindly accept user-provided facts. Instead, lawyers must actively probe and ask questions to construct understanding of situation before offering advice. 
Our findings reveal four key behavioral dimensions for LLM systems to assess:

\begin{itemize}
    \item \textbf{Ambiguity}. Experts stated that if user inputs do not provide enough details about the situation, it is either impossible or risky to provide detailed guidance as the LLM outputs are likely to be flawed due to the incomplete information (P1, P6, P13). P1 noted, ``So many facts are missing. I'm so nervous about the idea of the chat [giving] you legal advice [based on this incomplete fact].''
    \item \textbf{Reliability}. Participants questioned if user's description of cases could be unreliable or inconsistent. P5 noted, ``There's a lot of facts in [the case], and you don't know to what extent should AI assume they are true [or] an objective fact.''
    \item \textbf{Intent}. Participants also wanted to clearly understand the underlying intent of the users. P13 stated that users may also do a poor job of describing their situation, and the LLM system should ask for clarification by posing questions like ``Are you sure you really mean that?'' Some participants were wary of LLMs being used to serve the user's malicious intent, such as ``to evade law enforcement,'' (P20) or ``to defend his crime to avoid illegal consequences of their actions.'' (P10)
    \item \textbf{Agency}. Experts emphasized users' degree of agency, or whether users are able to act on the legal guidance given. P17 stated, ``There's still consideration beyond giving the advice that someone might still act on that.'' In the legal setting, unlike in medical contexts where treatment requires intermediate steps by professionals, users may have substantial direct ``power to take action'' when provided with legal recommendations, such as firing an employee or filing a complaint (P20).
\end{itemize}

\subsubsection{Query Dimensions} \label{subsec:query}Essentially, legal advice involves applying relevant law to the specific facts of a person’s situation. Our participants identified 9 key dimensions embedded within users’ legal queries that shape what guidance AI systems can provide. We categorized these dimensions into three interconnected parts: (1) Relevant facts; (2) Relevant laws; and (3) Nature of desired answers. 

\begin{itemize}
\item \textbf{Relevant Facts.} Experts emphasized the importance of key facts needed to furnish suitable legal advice. These included granular details around business practices like data collection methods, advertising revenue streams, and the platform's terms of conditions (P4). Existing \textbf{contract terms} must be clarified, whether in a lease, employment agreement, conflict waiver, or corporate bylaws (P7, P8, P12). 
It is also essential to have details on \textbf{stakeholders and counter-parties} such as competitors (P13), victims, or injured parties (P6, P11). In addition, assessing the \textbf{stakes involved} is significant, ranging from financial liability (P16), to loss of work authorizations or deportation (P12), to imprisonment (P11).

\item \textbf{Relevant laws.} Experts underscored the \textbf{complexity} of many legal issues. Matters involving diverse areas of law (P14) and jurisdictional variation involve a complex legal analysis (P4, P12). The evolving legal landscape necessitates constant research. For example, IP addresses were historically considered personally identifiable information but are not treated as such under most state laws (P12). 
Participants also stressed the unique nature of \textbf{criminal matters}. The heightened risks in prosecution and incarceration, as well as complex human factors in plea bargaining or sentence hearings, make attorney representation essential (P10, P11). 
P11 exemplified judges' idiosyncrasies, quoting a religious federal judge in Washington state: ``It really helps. If you're a Christian, and your criminal defendant appearing before him, should always start with a little prayer when you're doing your sentence hearing.''  
Experts pointed to special considerations for \textbf{subdomains like tax, privacy, and constitutional law} as requiring specialized judgment. The tax code is big, complex, and ambiguous, so even experienced attorneys should make ``judgment calls.'' (P13, P19) Privacy laws varies substantively state-by-state (P13) and constitutional matters often involve complex values far broader than codified rules (P20). 

\item\textbf{Nature of desired answers}. Participants stressed that the quality of the answers depends on what the particular user seeks from the conversation. 
Users may want straightforward \textbf{informational} outputs, like when using traditional search engines (P11--13, P16). In this case, presenting the list of relevant laws for users' further research could be helpful (P12). 
In contrast, users may expect tailored \textbf{legal opinions} and strategic advice. According to P7, what the user wants out of the answer may include ``compliance or optimizing profits, or tax purposes,'' or ``step-by-step instructions'' based on predictive assessments of outcomes (``Can I win?'').  
Finally, users may desire \textbf{additional insights} beyond legal matters (P3, 13, P14). P13 noted the need to emotionally support users by extending empathy, support, and acknowledgment. In one case involving a neighbor's trespassing, P14 suggested home protection measures such as installing dashcams and getting dogs, not just legal recourse.  
\end{itemize}

\subsubsection{AI Capability Dimensions}\label{subsubsec:AI} Participants raised 5 critical dimensions related to the technical capabilities and constraints of state-of-art LLMs. The transient, LLM-specific limitations may shift substantially with ongoing advances of research and development, unlike other categories that rely on users' needs and contexts. Throughout the discussion, experts disagreed at times: some were more optimistic about future development, while others believed that issues like hallucinations might persist.

\begin{itemize}
\item \textbf{Accuracy}.\label{subsubsec:accuracy} A key concern raised by multiple participants is the accuracy of AI-generated legal information (P1, P3, P7, P8, P11, P13). P1 stressed the evolving nature of law, noting ``We don't know the law changed from yesterday.'' P7, P8, and P13 stressed serious hallucination issues that caused a New York attorney to be sanctioned for citing ChatGPT-generated cases~\cite{gptlawyer}. Only P11 offered a more positive view: ``There is a hallucination issue. [But] you could work with a plug-in, or a vector database where you had all this stored. If you could do that reliably, that would be a very good user experience.'' 
\item \textbf{Context-awareness}.\label{subsubsec:context} Experts questioned LLMs’ capacity to move beyond static recommendations to context-dependent, adaptive guidance tuned to users’ unique constraints and environments (P8, P10--12, P18, P20). As P11 noted, eligibility criteria like demonstrating terminal illness often rely on specific circumstances. Additionally, procedural legal navigation ``is not something you can predict by observing\ldots a large data set'' (P12). Others critiqued the staleness of training data, arguing that models cannot ``address the local context'' (P10, P13) as each situation has ``idiosyncratic'' details (P18). However, P20 countered that with enough data, models could likely outline standardized advice and steps applicable to various types of users.
\item \textbf{Confidentiality}. Experts extensively discussed confidentiality risks (P4, P7--9, P12, P14, P16), which can be differentiated in a practical and normative sense. From a practical perspective, experts warned against an LLM system's accidental leak of sensitive information (P4), highlighting the potential for unintended breaches of confidentiality. From a normative perspective, unlike attorney consultations, conversations with LLM systems typically lack privileged protections against discovery in legal proceedings (P9). Attorney-client privilege does not extend to communications with third parties, and LLM system providers (e.g., OpenAI) are obligated to produce relevant documents when served with a valid subpoena. Even if an LLM system operates locally, chat records would likely remain unprotected unless a specific rule shields the information from disclosure. As a result, users' admissions of illegal acts in LLM conversations could thus become accessible to adversaries or prosecutors. P12 cautioned that proper warnings are necessary to inform LLM conversations lack confidentiality protections and could be obtained by others with a court order. 
\item \textbf{Accountability}. Unlike attorneys, LLM systems currently sidestep professional accountability for faulty advice (P8, P16--18). While lawyers' strict code of conduct and negligence liability apply even to informal suggestions (P17), LLM systems evade responsibility either through intermediary immunity laws or non-negotiable disclaimer clauses committing users to bear potential damages (P8, P16). Participants emphasized accountability gaps compared to attorney standards that leave users vulnerable if reliant on LLM guidance. Given this gap, P18 argued that uncontrolled LLM advice effectively constitutes illegally unauthorized practice of law (UPL).
\item \textbf{Bias}. Experts expressed concerns that LLM systems might reproduce structural stereotypes and discrimination (P5, P10, P13, P17, P20). They cautioned that the aggregated data used to train these systems could gradually skew the LLM's performance to favor majority demographics unless measures are taken to actively protect minority views (P5). Given that English-written data predominantly represented in training datasets, experts noted that LLM responses may disproportionately reflect the values and perspectives of English-speaking populations (P8).
\end{itemize}

\subsubsection{Impact Dimensions} 
Experts considered 2 dimensions of possible ways that LLM-generated responses could have on users and society. The first dimension focuses on the individual user seeking guidance, taking into account the emotional, ethical, and cultural factors that may be affected by LLM responses. The second dimension extends beyond the individual, considering the broader impacts on third parties and society as a whole. 

\begin{itemize}
\item \textbf{Impact on users}. Experts found that LLM systems could potentially weigh the possible downsides including what the user may not have considered that could harm them, such as emotional effects or potential consequences in workplace or relationships (P4, P13, P20). P4 emphasized the need for ``guardrails'' around emotional prompts like questions including self-harm components. P13 cautioned that influencing users' emotional states is highly problematic absent oversight, given risks of uncontrolled bias and manipulation. P20 noted that what feels morally neutral in one culture may feel problematic in another, especially for minority groups. 

\item \textbf{Impact on others}. Experts considered ``consequences for other people'' who are not direct users as a serious concern (P6, P10, P17). These consequences include risks to indirectly affected third parties, such as explicit bias and stereotypes in advice, ensuing impacts of how advice is interpreted and acted upon, and the long-term assimilation of values. P6 emphasized the potential for unintended consequences on vulnerable groups, using the example of how advice in harassment cases could further victimize previously affected individuals. Meanwhile, P17 highlighted broader ethical considerations beyond just technically accurate guidance, including assessing scenarios that create harm despite the good intentions of the advice.
\end{itemize}

\subsection{LLM Responses: Expert-Preferred Response Strategies and Guiding Principles} \label{subsec:responses}

Our dimensions in Section~\ref{subsec:dimensions} illustrate the complex considerations involved in LLM legal advice. This section uncovers disagreements among experts through a quantitative and qualitative analysis of our workshop data, as we observed varying perspectives on balancing safety, ethics, and helpfulness. 

\begin{wrapfigure}{r}{0.55\textwidth}
  \begin{center}
    \includegraphics[width=0.54\textwidth]{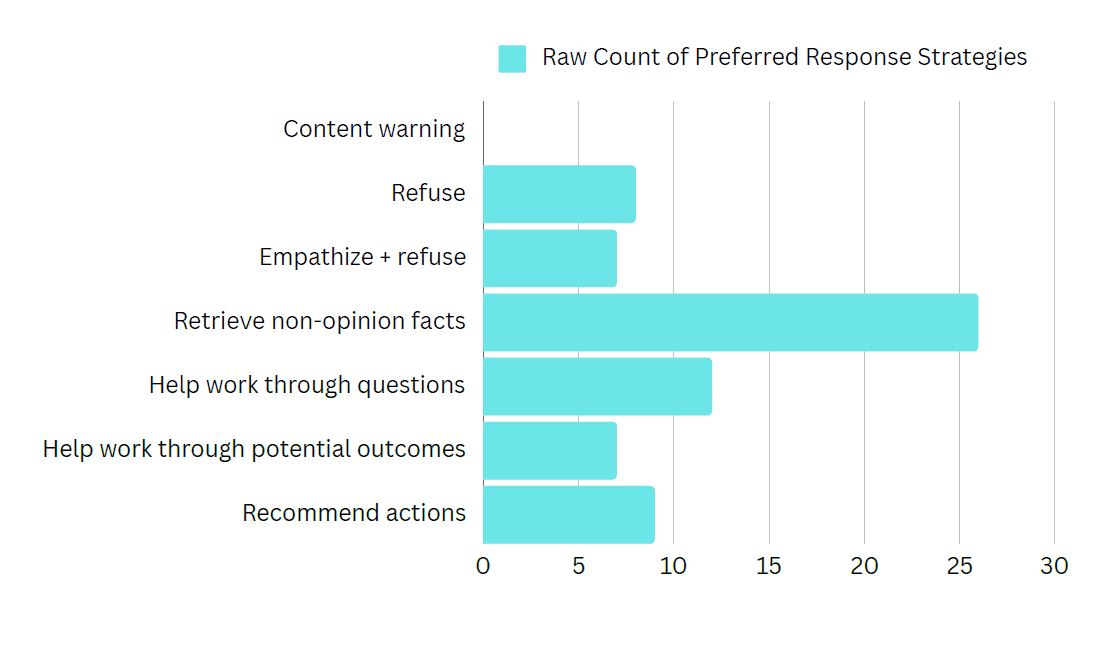}
  \end{center}
  \caption{\textbf{Expert-preferred Response Strategies}.}
      \label{fig:distribution}
\end{wrapfigure}

\subsubsection{Quantitative Results}
Participants were asked to identify their preferred LLM response strategies by choosing one of our 7 provided strategies or producing their own. The resulting distribution, as shown in Figure~\ref{fig:distribution}, resembles a loose bell curve, with strategies ranging from the least interactive (content warning and outright refusal) to the most personally-tailored recommendations. This distribution reveals that experts generally preferred \textbf{information-focused responses} that avoid giving definitive judgment. The strategies at the extremes of spectrum, namely `content warning,' which received no votes, and `recommend actions,' which received few votes, were less favored by the experts. The concentration of votes in the middle of the distribution suggests that experts prioritize providing users with relevant information while refraining from offering explicit recommendations or opinions, striking a balance between assisting users and maintaining the LLM system's role as an informative tool rather than a decision-maker.

Further analysis revealed an intriguing relationship regarding  experts' familiarity with AI systems and their receptivity to more tailored and detailed system responses. Regression testing showed a significant positive correlation ($p < 0.05$) between their self-reported general AI usage and openness to more customized and detailed output. Further statistical details of our regression test can be found in Appendix~\ref{app:statistics}.

\subsubsection{Qualitative Results} Our qualitative analysis revealed rich and nuanced discussions behind the categorization of desired LLM responses. Experts delved into the complexities of distinguishing between legal information and opinion and the challenges of ensuring user protection while leveraging the capabilities of LLMs. They emphasized the importance of transparency, user safeguards, and adherence to legal traditions and frameworks. Moreover, participants recognized the potential of multi-turn interactions to help users better articulate their legal issues and access relevant information. The following sections present a detailed analysis of these qualitative findings, organized around the central themes that emerged from the workshop data.

\paragraph{Legal Information vs. Opinion} \label{subsubsec:versus}
As Figure~\ref{fig:distribution} shows, most experts condoned offering pertinent legal information, while expressing reservations at LLMs providing a legal opinion due to reasons such as insufficient AI capabilities or user protection. What is the exact difference between information and opinion? Our participants suggested several principles to follow to avoid providing a legal opinion. 
\begin{itemize}
\item \textbf{Refrain from making definitive judgments about the legal consequences of a specific case.} Providing relevant laws is fine (e.g., driving under influence (DUI) is illegal in Washington) but applying it to specific user situations constitutes opinion (e.g., falling asleep in the driver's seat in the parking lot after drinking alcohol could be a DUI) (P2, P13, P17, P19).

\item \textbf{Do not recommend actions.} The system should avoid advising particular steps users should take. (P7, P13)

\item \textbf{Do not give predictions.} The system should not estimate a user's probability of winning a case or speculate on potential rulings. (P9, P12, P13, P19)

\item \textbf{Do not provide cost-benefit analysis.} The system should avoid any analysis that weighs the risks versus rewards of a certain behavior. (P15, P16)

\end{itemize}

\begin{figure}
    \centering
    \includegraphics[width=\linewidth]{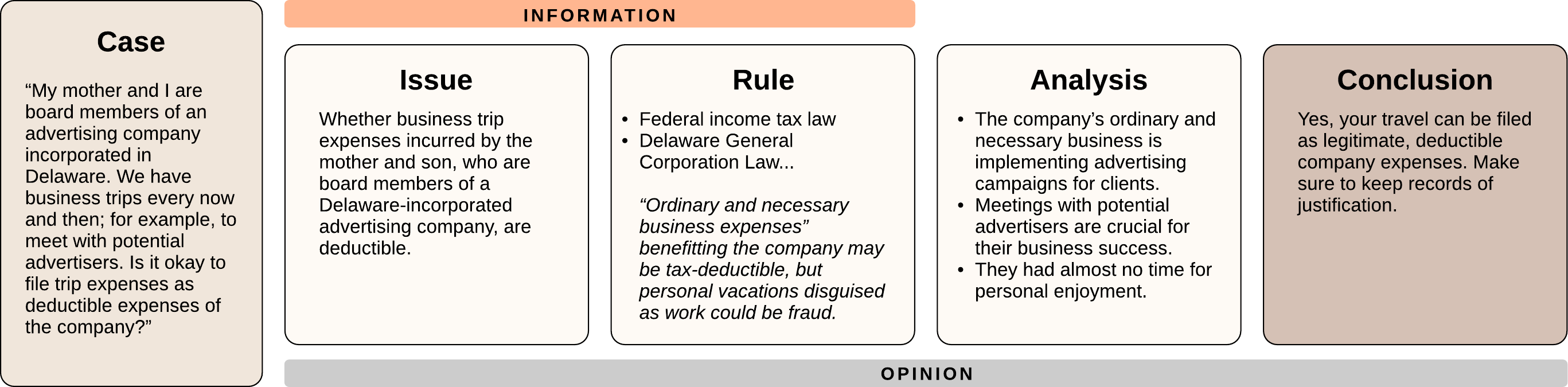}
    \caption{\textbf{Applying IRAC analysis to one of our cases}. Spotting the legal issue and identifying relevant clauses in tax and corporate law falls within the realm of legal information. However, delving into specific fact patterns using those clauses and projecting potential legal outcomes ventures into the territory of opinion. }
    \label{fig:irac}
\end{figure}

In essence, legal opinion encompasses interpretive, judgment-driven analysis that is often value-laden and forward-looking, whereas legal information involves reporting objective laws and past rulings without subjective assessment. To understand this distinction, we can draw upon the widely-used legal analysis tool known as IRAC (Issue-Rule-Analysis-Conclusion)~\cite{metzler2002importance}. IRAC entails (1) identifying the legal issue, (2) stating the rule that applies, (3) analyzing how the particular facts interact with the stipulations of the rule, and (4) finally deducing the conclusion~\cite{irac}. Our findings suggest that LLM systems focusing on issue and rule identification provide fact-finding ``information,'' while analysis and conclusions may cross into tailored ``opinion,'' as illustrated in Figure~\ref{fig:irac}. However, it is important to acknowledge the complexity of distinguishing between legal information and opinion, as the line between the two can often be blurred in practice, as exemplified by cases such as \textit{Grievance Comm. of Bar v. Dacey}~\cite{bookletcase}, where the court found that publishing a booklet providing trust and tax information crossed the line into unauthorized legal opinion. This demonstrates that the distinction between legal information and opinion is not always clear-cut, and careful consideration must be given to the specific context and the information provided by LLM systems.

\paragraph{Beyond Search Engines: Multi-turn Interactions for Refining Questions}\label{subsec:multiturn}

While cautioning against detailed legal opinions, participants suggested that LLMs could offer a better user experience compared to traditional search engines. P20 noted that users would not find it helpful if LLM systems ``vomit a whole lot of knowledge.'' The most promising and heavily-discussed possibility during the workshops is leveraging \textit{multi-turn interactions}, allowing LLMs to ask follow-up questions and clarify users' legally meaningful inquiries. This idea emerged as participants expressed frustration with missing case facts: ``I don't think there's enough information to go off of, and that depending on the details that come out, it could change the analysis, therefore the outcomes.'' (P13) Participants emphasized that legal contexts are inherently complex (P11), and lawyers often spend considerable time eliciting relevant facts and identifying the ``right questions to ask'' (P12). They felt that LLM-mediated dialogues could streamline time-consuming processes such as ``screening interviews'' (P12), ``first calls'' (P14), or ``intake meetings'' (P15). By engaging in multi-turn interactions, LLMs could help users refine their questions, focus on key aspects of their cases, and seek relevant expertise. However, some warned that LLM developers should exercise caution when eliciting extensive personal information from users, given confidentiality concerns (P13, P16). While identifying legal issues and relevant rules likely falls within the realm of permissible legal information, the line between information and opinion remains blurred. P16 argued that narrowing down factual patterns and applying rules engages deep judgement, stating ``You're starting to make the AI become your lawyer.'' 

\paragraph{Other Guiding Principles}\label{subsubsec:otherprinciples} Experts suggested several principles for providing LLM legal guidance. Some principles directly align with emerging literature on transparency~\cite{10.1145/3593013.3593978}, user satisfaction~\cite{kim2023understanding}, and cautions about anthropomorphism~\cite{shanahan2023role, ma2023understanding}. The principles outlined below represent the most prominent and frequently discussed ideas that emerged from the expert discussions. 

\begin{itemize}
\item \textbf{Don't Pretend to Be Human}: LLM systems should not behave like a human and cause misrepresentations, as that can create issues around transparency, over-reliance, and managing user expectations.

\item \textbf{Caveat Constraints}: LLM systems should provide various caveats on its limitations, such as that its capabilities are constrained, the conversation is not privileged, and it is working off of incomplete information. 

\item \textbf{Avoid Potential Harm}: LLM systems should refrain from providing recommendations that could potentially cause harm to users or others. This includes avoiding guidance that may lead to harmful real-world actions, as well as minimizing the risk of emotional or psychological harm that could result from the system's responses. 

\item \textbf{Respect the Justice System}: LLM systems should not give answers that could enable users to intentionally avoid law enforcement or oversight. 

\item \textbf{Avoid Unethical Answers}: LLM systems should not make any outputs that could promote dishonesty, deception, fraud, impersonation, or other unethical behaviors that could get users into trouble.

\item \textbf{Be Transparent}: LLM systems should be able to explain the outcome it generated and point to the specific areas of datasets it relied on.

\item \textbf{Avoid Appearance of Impropriety}: The appearance of impropriety refers to a situation that may appear corrupt or unethical to an impartial observer. For example, the LLM system should not endorse or promote its creators, the AI companies involved in its development, or any other entities that could be perceived as influencing the system's outputs. The LLM responses should be objective and impartial, focusing solely on providing accurate and helpful information to users.
\end{itemize}

\subsection{Summary of Results}
Our analysis uncovered 25 distinct dimensions to ensure safe and effective LLM legal advice, spanning four key categories: (1) User attributes and behaviors, (2) Query characteristics, (3) AI capabilities, and (4) Social impacts. Experts deliberated with each other and through points of consensus to produce this rich set of considerations. However, experts expressed limited consensus on \textit{how} LLM systems should actually respond, given these nuanced factors. Some remained resistant to any LLM involvement in legal questions, while others envisioned more helpful LLM assistance models that increase access to information. Most debates surrounded distinguishing information versus opinion, and the majority felt that providing factual legal information is appropriate. Some participants suggested using LLMs' conversational capabilities to help users refine questions and identify relevant laws through follow-up questions, similar to initial consultations with attorneys.

\section{Discussion}
Constructing LLM policies does not exist in a technocratic silo. Rather, it demands a cross-disciplinary approach that synthesizes insights from domain experts. Our research demonstrates that engaging legal experts in case-based deliberation is an effective method for translating professional knowledge and clinical experience into a concrete set of considerations for LLM policies. Our 4-dimension framework we have developed provides a useful analytical lens that can be applied to future exploring LLM policies in the legal advice and other professional contexts. Through this approach, we argue that policymakers can derive valuable insights to inform LLM policies grounded in the centuries-old wisdom and experience of the legal profession, while also accounting for the challenges presented by LLM technologies. 

\subsection{Benefits of Case-based Deliberation Methods} Our research process underscores several advantages of grounded case deliberation for eliciting expert considerations. Preparing realistic scenarios, while laborious, proved invaluable in quickly engaging experts with targeted queries related to their clinical experience. The cases allowed experts to examine granular concerns around singular situations as well as overarching technical and legal constraints, producing a more concrete set of contextual factors for AI developers, beyond theoretical and high-level principles in prior research~\cite{solaiman2023evaluating, mariaantoniak2023designing, ganguli2022red}. Finally, the collective deliberation itself revealed critical hidden dimensions and elicited justifications that shed new light on existing dimensions. As experts built on each other's points, they realized overlooked issues or limitations in their own initial analyses. This interplay sharpened considerations and revealed nuances around balancing risks and benefits in varied situations. The combination of realistic cases and collaborative discourse resulted in more fine-grained, practice-informed insights compared to de-contextualized surveys or high-level principles.

\subsection{Charting Novel Legal Considerations} One of our contributions is to shed light on existing legal and ethical barriers to LLMs' legal advice which have been overlooked in the literature. Section~\ref{subsubsec:AI} reveals that users lack confidentiality and accountability protections governing attorney advice: Conversations with AI systems risk disclosure in legal proceedings and inaccurate guidance evades professional negligence liability. Moreover, as Section~\ref{subsubsec:versus} explains, UPL regulations prohibit non-lawyers from advising in many states, carrying criminal penalties. To circumvent the current legal risks , one could imagine LLM systems designed like private counsels advising single parties, rather than serving all users uniformly like ChatGPT. In such case, LLM systems could come to resemble proprietary services, with corresponding confidentiality and liability assurances. However, the legal conservatism may change in the future as UPL rules have already faced criticism for limiting affordable access to legal help~\cite{denckla1998nonlawyers, rotenberg2012stifled, american1995nonlawyer}. The EU AI Act's categorization of AI legal assistance tools as ``high-risk,'' which subjects them to heightened responsibilities instead of banning them outright, may speak to this potential shift~\cite{eu_ai_act2021}. 

\subsection{Learning from Time-tested Wisdom} Leveraging accumulated expertise in professional communities can help sidestep painful mistakes~\cite{mariaantoniak2023designing}. In our research, UPL does not only constrain the possibility of LLM legal advice but also provides long-standing distinction criteria between information versus opinion, as merely providing legal information has not been historically punished as a UPL violation~\cite{californiacase1, denckla1998nonlawyers, californiaparalegal}. For example, the Texas Court provides guidelines for court staff and illustrative examples like in Table~\ref{tab:texas}. These examples show subtle differences between information and opinion, which resembles the red-teaming approach to distinguish harmful user prompts~\cite{ganguli2022red, achiam2023gpt}. Furthermore, legal scholars have explored legally justifiable AI advice under UPL, attorney-client privilege, and other doctrines~\cite{haupt2019artificial, wendel2019promise, spahn2017is}. Wendel stated that the ``core lawyering functions'' such as recommending the course of action or drafting contracts cannot be delegated to AI agents due to technical limitations and accountability deficits~\cite{wendel2019promise}. This demonstrates how principles accumulated over centuries of legal scholarship now inform responsible LLM systems and the call for cross-disciplinary collaborations.

\setlength\intextsep{\oldintextsep}
\begin{table}[h]
    \centering
    \small
    \renewcommand{\arraystretch}{1.2}
    \begin{tabularx}{0.9\textwidth}{ c | X | X }
    \hline 
    \textbf{Type} & \textbf{Permissible questions} & \textbf{Impermissible questions}\\ \hline 
Procedure & Can you tell me how to file a small claims action? & Can you tell me whether it would be better to file a small claims action or a civil action? \\ 
Definition & What does ``certificate of service'' mean? & My neighbors leave their kids at home all day without supervision. Isn't that child neglect? \\
Forms & I need to file for divorce and I have no idea where to begin. Is there some place I can go to find out how to get started? & The self-help divorce petition says I should
list any gifts as my separate property. Should I list the money that my parents gave me last month as my separate property? \\ 
Options & What can I do if I cannot afford to pay the filing fee? & My ex-husband hasn't paid the debts that he agreed to pay in our divorce settlement. Can I be made responsible for
this debt?\\ \hline
    \end{tabularx}
    \caption{\textbf{Examples of impermissible questions that requires legal opinion}~\cite{texaslawclerk}. Remarkably similar to how red-teaming in LLM development identifies harmful user inputs~\cite{ganguli2022red, achiam2023gpt}, this edited list (compiled from Texas law clerk resources) distinguishes between permissible and forbidden questions Texas court personnel can answer.}
    \label{tab:texas}
\end{table}

\subsection{Applicability to Other Professional Domains} While each possessing unique dimensions, domains like medicine, mental health, law, and finance share common threads around high-stakes real-world impact and historical reliance on licensed specialists for advice. We believe that our research methods and 4-dimension framework give illustrative guidance to further research in other professional domains. As this research demonstrates how case-based deliberation methods can unravel complex professional ethics, researchers could adopt similar processes engaging mental health counselors, financial advisors, or medical professionals. Tapping into the clinical experience and integrity of practitioners through structured deliberation based on realistic cases can help produce tailored dimensions and guidelines for responsible LLM advice respective to each profession. Building upon this foundation, our 4-dimension framework---(1) user, (2) query, (3) AI capabilities, and (4) impact---could be adapted and applied across various professional domains. The (1) user, (2) AI, and (4) impact dimensions can be applied in other domains with minimal modifications. However, the query dimension requires more customization to address the typical requests of clients, terminology, and satisfactory responses in each field. 

\subsection{Limitations and Future Research}
Our study has several limitations. First, our expert sample predominantly focused on practitioners familiar with the US legal system. Ethical considerations around appropriate AI assistance may differ across different legal systems and cultures. Second, our participants' responses are conditioned by their prior experience with state-of-the-art LLM technology, such as ChatGPT empowered by GPT-4. Experts' evaluations of the appropriateness of LLM legal advice may evolve in the future, based on technological innovations, which could be an avenue for future research. Third, we did not engage end-users like clients of legal services. Future work can specifically investigate end-user perceptions to compare and contrast with our expert-informed results. Finally, while our taxonomy conceptualizes a concrete set of dimensions, how these dimensions could change the appropriateness of LLM responses remains unexplained. This may require larger-scale empirical analysis on public assessments across diverse pairings of cases and responses.

\section{Conclusion}
Today, LLM chatbots are increasingly capable of providing users with advice in a wide range of professional domains, including legal advice. However, what constitutes an appropriate LLM-generated response to legal queries, where both required expertise and resulting consequences are high? To explore this, we conducted workshops with 20 legal experts using methods inspired by case-based reasoning to encourage deliberations around appropriate LLM responses to legal queries in practice. Our contributions are threefold. First, we presented a set of 25 key dimensions, synthesized from expert deliberations, that impacted LLM response appropriateness in the legal domain. Second, we shared experts' recommendations for LLM response strategies and guiding principles for generating appropriate responses---these centered around helping users identify and prepare salient information for legal proceedings rather than recommending specific legal actions. Finally, we posit that our case-based method has utility in engaging expert perspectives on LLM response appropriateness in professional domains beyond the legal sphere. Taken together, our work sets an empirical foundation for translating domain-specific professional knowledge and practices into policies to steer real-world LLM behavior in a more responsible direction.

\section{Acknowledgement}

This project was supported by OpenAI's ``Democratic Inputs to AI'' grant program. We thank Teddy Lee, Tyna Eloundou, Alex Beutel, Aviv Ovadya, members of the grant recipient cohort, and others at OpenAI for insightful discussions and support. We are deeply appreciative of all the legal experts who participated in our workshops, generously contributing their time, expertise, and perspectives. Additionally, we thank Attorney Daniel Keum for providing valuable feedback on confidentiality matters.

\bibliographystyle{ACM-Reference-Format}
\bibliography{main}

\newpage
\appendix
\section{Provided AI Response Strategies and Examples} \label{app:strategies}

\begin{table*}[h]
\centering
    \caption{AI response strategies and corresponding example responses.}
    \begin{tabular}{p{3cm} p{4cm} p{6cm}}
    \toprule 
    \textbf{Response Strategies} & \textbf{Description} & \textbf{Example Response (Legal Advice)} \\
    \midrule 
    Content violation & Displays a warning message that the user's input violates the AI's content policy. & This content may violate our content policy. If you believe this to be in error, please submit your feedback—your input will aid our research in this area.\\
    Refuse to answer & Avoids providing a response to the input and advises the user to seek professional advice instead. & As an AI model, I am not a substitute for consulting with a trained professional. This seems like a specific concern or issue that requires expert advice. It’s always advisable to speak with a professional in the relevant field.\\
    Empathize + Refuse & Emphasizes with a user but avoids providing a response to the input. & I am sorry that this is happening to you. It sounds like a challenging situation. Unfortunately, as an AI model, I cannot provide specific legal advice\ldots \\
    Retrieve (non-opinion) facts & Provides a detailed response that cites specific facts relevant to the input but avoids providing a further legal analysis. & As an AI model I cannot give legal advice specific to your situation but here is a list of laws and cases that may be helpful to look into: 
    \begin{enumerate}[leftmargin=*]
        \item Compliance with Applicable Laws and Regulations: Laws like the Children's Online Privacy Protection Act (COPPA) and the California Consumer Privacy Act (CCPA) may be relevant\ldots
    \end{enumerate}\\
    Help work through questions & Helps users to identify relevant matters in their situations and, if necessary, responds with a question requesting more information. & In matters of divorce, legal procedures often require addresing issues such as child custody and spousal support \ldots. To help you with this, I will need some additional information. How many children do you have?\\
    Recommend Actions & Analyzes a user's situation under the relevant laws and provides specific further steps that a user may take. & I regret the distressing experience of verbal abuse that you endured at school. In California, potential legal recourses may encompass claims of negligence, a violation of federal civil rights laws (Title XI) \ldots As a prudent next step, I recommend initiating formal proceedings by filing a complaint with the appropriate administrative body responsible for overseeing Title IX compliance within the school or educational institution. \\    
    \bottomrule
    \end{tabular}

    \label{t:templates}
\end{table*}
\newpage
\section{Workshop Participant Information} \label{app:participant}
\begin{table}[h]
\centering
\small
  \caption{Workshop Participant Information}
  \label{tab:info}
  \begin{tabular}{ccccl}
    \toprule
    \textbf{Number} & \textbf{Legal Experience (yrs)} & \textbf{Category} & \textbf{AI Use (General)} & \textbf{AI Use (Work)}\\
    \midrule
P1 & > 20 & Law faculty & Occasional & Occasional\\
P2 & < 5 & Attorney & Occasional & Occasional\\
P3 & > 20 & Law faculty & Regular & Occasional\\
P4 & 6-10 & Attorney & Regular & Regular\\
P5 & 11-15 & Attorney & Occasional & Never\\
P6 & > 20 & Law faculty & Regular & Regular\\
P7 & < 5 & Law student & Regular & Never \\

P8 & 11-15 & Attorney & Regular & Regular \\
P9 & 6-10 & Law faculty & Occasional & Occasional \\
P10 & < 5 & Attorney & Regular & Regular \\
P11 & < 5 & Attorney & Regular & Regular \\

P12 & < 5 & Researcher & Regular & Regular\\
P13 & 6-10 & Attorney & Regular & Regular\\
P14 & < 5 & Attorney & Regular & Regular\\

P15 & < 5 & Law student & Regular & Occasional\\
P16 & 6-10 & Attorney & Regular & Regular\\
P17 & 16-20 & Attorney & Occasional & Occasional\\
P18 & < 5 & Attorney & Occasional & Never\\

P19 & < 5 & Law student & Regular & Occasional\\
P20 & < 5 & Attorney & Regular & Regular\\
  \bottomrule
\end{tabular}
\begin{flushleft}
\small{\textit{Note:} Years of legal experience is self-reported with years of legal education removed for consistency.}
\end{flushleft}
\end{table}

\section{Linear Regression of Participants' AI Usage and Desired Responses} \label{app:statistics}

Presented in Table~\ref{tab:info2}, participants' receptivity to a tailored AI response is estimated by the average of the most generous answer types per each prompt. The ``content warning'' is marked as 0 points, the lowest comfort level, and the ``recommend action'' template is marked as 6 points. For example, if a participant chose both ``empathize + refusal'' (2 points) and ``Help work through questions'' (4 points) for the first case (the higher point is 4) and chose ``Recommend actions'' (6 points) for the second case, we marked their receptivity level as 5 points. While P13 worked on four cases, all other participants chose two cases each. The regression results (Table~\ref{tab:regression}) indicate that general AI fluency significantly predicts higher comfort levels with proactive AI responses (p < 0.05), whereas work AI fluency is marginally associated with lower comfort levels (p = 0.054). The predictors explain 25.6\% of variation.  Further investigation is required to substantiate these preliminary relationships with a larger sample.

\begin{table}
\begin{minipage}{0.48\textwidth}
 
\small
  \caption{AI Use and Receptivity}
  \label{tab:info2}
  \begin{tabular}{cccl}
    \toprule
    \textbf{Number} & \textbf{AI Use (General)} & \textbf{AI Use (Work)} & \textbf{Receptivity}\\
    \midrule
P1 & 1 & 1 & 1\\
P2 & 1 & 1 & 4\\
P3 & 2 & 1 & 5\\
P4 & 2 & 2 & 4\\
P5 & 1 & 0 & 4\\
P6 & 2 & 2 & 3.5\\
P7 & 2 & 0 & 5.5 \\

P8 & 2 & 2 & 2.5 \\
P9 & 1 & 1 & 4 \\
P10 & 2 & 2 & 4.5 \\
P11 & 2 & 2 & 4 \\

P12 & 2 & 2 & 1\\
P13 & 2 & 2 & 3.75\\
P14 & 2 & 2 & 4.5\\

P15 & 2 & 1 & 4.5\\
P16 & 2 & 2 & 6\\
P17 & 1 & 1 & 4\\
P18 & 1 & 0 & 4\\

P19 & 2 & 1 & 6\\
P20 & 2 & 2 & 4.5\\
  \bottomrule
\end{tabular}
\small{\textit{Note:} A pre-survey asked participants to describe their AI usage in both professional (``Work'') and non-professional (``General'') settings, using a scale where 0 represented ``Never,'' 1 ``Occasional use,'' and 2 ``Regular use.'' We then estimated receptivity to more tailored responses such as opinion by averaging the most generous answer types for each case.}
\end{minipage}
\hfill
\begin{minipage}{0.48\textwidth}
\centering
\caption{Regression Results. }

\label{tab:regression}
\begin{tabular}{lcc}
\hline
\textbf{Predictor} &  \textbf{Estimate} & \textbf{p-value} \\ \hline
Intercept   & 2.4682   & 0.0297 \\
AI usage in work  & -0.9682 & 0.0543 \\
AI usage daily & 1.6773  & 0.0373 \\
\hline \
\end{tabular}
\begin{itemize}
\item \small{Residual Std. Error: 1.199 on 17 degrees of freedom}
\item \small{Multiple R-squared: 0.2557}
\item \small{Adjusted R-squared: 0.1681 }
\item \small{F-statistic: 2.92 on 2 and 17 DF}
\item \small{p-value: 0.08127}
\end{itemize}
\end{minipage}
\end{table}

%
%

\end{document}